\documentclass[journal ]{IEEEtran}
\usepackage{lineno}
\usepackage{amsmath,amssymb}
\usepackage{float}

\usepackage{booktabs}
\usepackage{makecell}
\usepackage{float}
\usepackage{multirow}
\usepackage{graphicx}
\usepackage{tabu,bm}
\usepackage{color}
\usepackage{subfigure}
\usepackage{booktabs}

\usepackage[linesnumbered,ruled,vlined]{algorithm2e}
\usepackage{algorithmic}
\usepackage{url}

\usepackage{cite}
\ifCLASSINFOpdf
\else
\fi
\begin{document}
%
\title{Task-Oriented Multi-User Semantic Communications for VQA Task}
%
%
%

\author{Huiqiang Xie,~\IEEEmembership{Student Member,~IEEE},
Zhijin Qin,~\IEEEmembership{Senior Member,~IEEE,} 
 and Geoffrey Ye Li,~\IEEEmembership{Fellow,~IEEE,}
\thanks{Huiqiang Xie and Zhijin Qin are with the School of Electronic Engineering and Computer Science, Queen Mary University of London, London E1 4NS, UK (e-mail: h.xie@qmul.ac.uk, z.qin@qmul.ac.uk).}
\thanks{Geoffrey Ye Li is with School of Electrical and Electronic Engineering, Imperial College London, London SW7 2AZ, UK (e-mail: geoffrey.li@imperial.ac.uk).
}
}


\maketitle

\begin{abstract}
Semantic communications focus on the transmission of semantic features. In this letter, we consider a task-oriented {multi-user} semantic communication system for multimodal data transmission.   Particularly, partial users transmit images while the others transmit texts to inquiry the information about the images. To exploit the correlation among the multimodal data from multiple users, we propose a deep neural network enabled semantic communication system, named MU-DeepSC, to execute the visual question answering (VQA) task as an example. Specifically, the transceiver for MU-DeepSC is designed and optimized jointly to capture the features from the correlated multimodal data for task-oriented transmission. Simulation results demonstrate that the proposed MU-DeepSC is more robust to channel variations than the traditional communication systems, especially in the low signal-to-noise (SNR) regime.

\end{abstract}


%
\IEEEpeerreviewmaketitle

\section{Introduction}
%
%
%
%
%
The continuously increasing number of connected-mobile devices and enriched  intelligent demands cause the explosion of wireless data traffic, which brings new challenges to communication systems, including  providing the cornerstone for various intelligent tasks,  exploiting the limited frequency resource, and dealing with the huge volumes of data. Semantic communication, which only transmits the related information, is a promising solution to address these challenges due to its great potential to reduce required resources for transmission significantly \cite{qin2019deep, shi2021new}.

{Traditional communication systems convert data into bits at the transmitter and require accurate bit recovery at the receiver, which depends on good channel conditions and high SNRs. Semantic communications  transmit and recover the meaning of the transmitted content directly and require no accurate bit recovery, thus, are more robust to the channels.} Inspired by the emerging deep learning (DL) technologies, some initial works on DL-enabled semantic communications focus on semantic recovery at the receiver for  text\cite{gold2018, 9398576, xie2020lite}, image \cite{bourtsoulatze2019deep}, and speech \cite{weng2021semantic}. How to exploit semantic information for specific tasks at the effectiveness level is another key area and few researchers pay attention to this area. There exist works that only focus on the {single-modal} data, i.e., image classification \cite{DBLPLee2019} and image retrieval \cite{JankowskiGM21}. However, in the practical communication scenarios,  the system is required to gather, transmit, and fuse multimodal data from {multiple users}. This motivates us to develop multimodal {multi-user} semantic communication systems.

Multimodal data refer to {describing} scenarios with different modalities. Typical examples of multimodal data include audio, video, images from electro-optical sensors, and text from radio frequency sensors. If a task needs more than one modality to perform, multimodal data are correlated in the context. Compared with the situations with {single-modal} data, multimodal data can provide more information for intelligent tasks, introduce new degrees of freedom, and improve performance of intelligent tasks \cite{LahatAJ15, gao2020survey}. The recent successful approaches for multimodal data fusion are mostly based on neural networks, and representative techniques include Deep Belief Net (DBN), Stacked Autoencoder (SAE), Convolutional Neural Network (CNN), Recurrent Neural Network (RNN) \cite{gao2020survey}. Multimodal semantic communications employ more than one users to serve only one multimodal intelligent task,  which is suitable for the emerging autonomous scenarios, i.e., autonomous checkout at retail stores. To build a {multi-user} semantic communication system for supporting multimodal data, we face two challenges: how to extract the proper semantic information from each user and how to build a model for multimodal semantic information fusion at the receiver.

\begin{figure*}[!t]
	\centering
	\includegraphics[width=140mm]{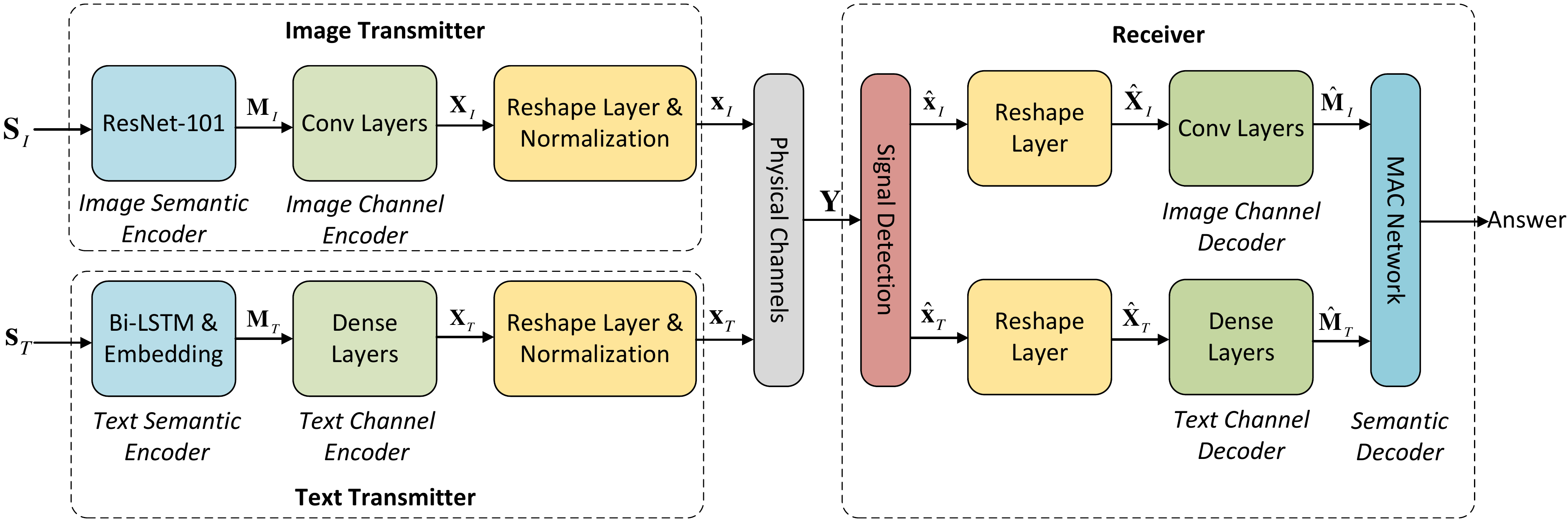}
	\caption{\textcolor{black}{The structure of proposed MU-DeepSC.}}
	\label{fig-2}
\end{figure*}

In this letter, we present our initial results in {multi-user} semantic communication for multimodal data. The detailed contributions are summarized as follows:
\begin{itemize}
\item A novel framework for the task-oriented multimodal data semantic communication system, named MU-DeepSC, is established, where the transceiver is jointly designed to perform the intelligent task. The visual question answering (VQA) task is adopted as an example to demonstrate the effectiveness of MU-DeepSC. 
\item The MU-DeepSC transmitter adopts the memory, attention, and composition (MAC) neural network to process the correlated data. By extracting the semantic information of {images} and text from different transmitters, the MU-DeepSC receiver will directly generate the answers based on the received semantic information at the receiver.
\item The simulations demonstrate that the proposed MU-DeepSC has the ability to transmit the image and text semantic information and perform data fusion at the receiver. 
\end{itemize}    
The rest of this letter is organized as follows.  Section II details the proposed MU-DeepSC. Numerical results are presented in Section III to show the performance of the DeepSC. Finally, Section IV provides the conclusion.

\section{Proposed  MU-DeepSC Transceiver}
In this section,  we adopt the image and text information as an example, where two users with single antenna and one receiver with $M$ antennas are considered for simplicity. It is easy to expand the network with the input of multi-image and multi-text. Besides, we design a deep neural network (DNN) for the considered semantic communication system to serve the VQA task, named as MU-DeepSC in Fig. \ref{fig-2}, of which the MAC network is adopted for answering questions. All models can be trained in the cloud and then broadcast to users.

\subsection{The Proposed MU-DeepSC}
 As shown in Fig. \ref{fig-2}, the proposed MU-DeepSC consists of an image transmitter, a text transmitter, and a receiver.

\subsubsection{Image Transmitter}
For the image transmitter in Fig.~\ref{fig-2}, which includes a semantic encoder and channel encoder. Particularly, the ResNet-101 is used for the semantic encoder and CNNs with different units are adopted for channel encoder to generate transmitted symbols. Before computing semantic information, images will be resized to a commonly used resolution, 224 $\times$ 224,  with bicubic interpolation, ${\bf S}_{\text{I}} \in \mathbb{R}^{1 \times 3\times224\times224}$. Then, we extract the semantic information by the image semantic encoder, which employs the first 30 blocks from the ResNet-101 network pre-trained on ImageNet.  Notice that the ResNet-101 model will be frozen during training,  as it has been well trained by more than one million images and is powerful enough to extract image semantic information. Then, the semantic image information can be extracted by the image semantic encoder, denoted as
\begin{equation}
    {\bf M}_I = {\cal SE}_{I}\left( {\bf S}_{I} ; {{\bm \alpha}_{I}}\right),
\end{equation}
where \textcolor{black}{${\bf M}_I \in {\mathbb R}^{1\times C_1 \times14\times14}$, where $C_1$ is the number of feature maps}, ${\bf S}_{I}$ is the input resized image and ${\bm \alpha}_{I}$ is the trainable parameters.

After passing through the semantic encoder, the captured semantic information is mapped to the transmitted symbols directly by the image channel encoder, which consists of CNN layers due to its characteristics of learning different local features of image effectively. Therefore, the transmitted symbols can be represented  by the image channel encoder, denoted by
\begin{equation}
    {\bf X}_I = {\cal CE}_{I}\left( {\bf M}_{I} ; {{\bm \beta}_{I}}\right),
\end{equation}
where ${\bf X}_I \in \mathbb{R}^{1\times C_2 \times 14 \times 14}$ is the compressed semantic image information, where $C_2$ is the compressed dimension with $C_2 < C_1$, and ${{\bm \beta}_{I}}$ is the trainable parameters.

In the above, ${\bf X}_I$ is real signal rather than complex, which is not suitable to transmission. Therefore,  ${\bf X}_I $ should be converted into the complex signal, ${\bf x}_I  \in \mathbb{C}^{1 \times 98C_2 }$, where $98C_2 = \frac{14\times14\times C_2}{2}$, firstly with reshape layer, which is then normalized by 
\begin{equation}\label{norm-1}
    l_{norm}({\bf x}_I) = \frac{{\bf x}_I}{\mathbb{E}( {\left\| {\bf x}_I \right\|_2 } )},
\end{equation}
for physical channel transmission.

\subsubsection{Text Transmitter}
In the text transmitter in Fig. \ref{fig-2},  the bi-directional long short term memory (Bi-LSTM) is used for the semantic encoder and dense layers with different units for the channel encoder to {generate} transmitted symbols. Assume that ${\bf s}_T=(s_{1, T}, s_{2, T}, \cdots, s_{L, T})$ is the transmitted sentence with $L$ words, where $s_{l, T}$ is the $l$-th word in the sentence to be transmitted. Before extracting its semantic information, the sentence will be embedded to map words to the numerical vector, ${\bf S}_T \in \mathbb{R}^{1\times L\times L_{embed}}$, by the embedding layer with $L_{embed}$ embedding dimension, which can be trained for better word representation.

We employ one layer Bi-LSTM to extract the semantic representations of  the input sentence. The corresponding text semantic encoder can be expressed as
\begin{equation}
    {\bf M}_T = {\cal SE}_{T}\left( {\bf S}_{T} ; {{\bm \alpha}_{T}}\right),
\end{equation}
where ${\bf M}_T \in {\mathbb R}^{1\times L \times K_1}$, where $K_1$ is the post-processed dimension from $L_{embed}$ to $K_1$, ${\bf S}_{T}$ is the embedded word vectors and ${{\bm \alpha}_{T}}$ is trainable parameters . Then, the text is compressed and mapped to the transmitted text symbols via the  channel encoder, which includes several dense layers to preserve all text semantic information to preserve the entire input information. Condescendingly, the transmitted symbols can be calculated by the image channel encoder as
\begin{equation}
    {\bf X}_T = {\cal CE}_{I}\left( {\bf M}_{T} ; {{\bm \beta}_{T}}\right),
\end{equation}
where ${\bf X}_T \in \mathbb{R}^{1\times L \times K_2}$ is the compressed semantic text information, where $K_2$ is the dimension after text channel encoder with $K_2 < K_1$, and  ${{\bm \beta}_{T}}$ is the trainable parameters.

Similar to the image transmitter, the transmitted signal, ${\bf X}_T$, will be {reshaped} into the complex signal, ${\bf x}_T \in \mathbb{C}^{1 \times \frac{K_2 L}{2}}$, firstly and normalized by
\begin{equation}\label{norm-2}
    l_{norm}({\bf x}_T) = \frac{{\bf x}_T}{\mathbb{E}( {\left\| {\bf x}_T \right\|_2 } )}.
\end{equation}

\subsubsection{Receiver}
The receiver is shown in Fig. \ref{fig-2}(c),  where convolution layers with different units are used for the image channel decoder, dense layers with different units for the text channel decoder, and the MAC network is adopted for the semantic decoder. The received symbols are detected firstly, then various semantic information is recovered through different channel decoders, and is finally merged the various semantic information to get answers. 

Assume that $V$ is the least common multiple between the length of image semantic and the length of text semantic information, the $M \times V$ signal received at the receiver can be expressed as
\begin{equation}
    {\bf Y} = {\bf HX} + {\bf N},
\end{equation}
where ${\bf H} \in {\mathbb{C} }^{M\times 2}$ is the channel between the BS and users,  ${\bf X}=\left [{\bf x}_{I},{\bf x}_{T} \right ]\in {\mathbb{C} }^{2\times V}$ denotes transmit symbols from text and image users in the considered system,  and $ {\bf N}\in {\mathbb{C} }^{M\times V}$ indicates the circular symmetric Gaussian noise, items of $\bf N$ are of variance ${\sigma }_{n}^{2}$. 

Here, we employ the additional domain knowledge, i.e., channel estimation, to improve the training speed and enhance the final decision accuracy. With the channel gain and zero-forcing detector, the transmitted signal can be estimated by 
\begin{equation}\label{eq13}
    \begin{aligned}
    {\bf{\hat X}} &= {\left( {{{\bf{H}}^H}{\bf{H}}} \right)^{ - 1}}{{\bf{H}}^H}{\bf{Y}}
     &= {\bf{X}} + {\bf{\hat N}},
     \end{aligned}
\end{equation}
where ${\bf{\hat X}}=[{\bf \hat x}_I; {\bf \hat x}_T]$ is the estimated information for the text and image users,  ${\bf{\hat N}} = {\left( {{{\bf{H}}^H}{\bf{H}}} \right)^{ - 1}}{{\bf{H}}^H}{\bf{N}}$ represents the impact of noise.  With operation in \eqref{eq13}, the channel effect is transferred from multiplicative noise to additive noise, which significantly reduces the learning burden. 

After signal detection, the estimated complex signals will be reshaped to the size  suitable for the following neural networks with reshape layer firstly,  ${\bf \hat x}_I \to {\bf \hat X}_I : \mathbb{C}^{1 \times 98C_2} \to \mathbb{R}^{1\times C_2 \times 14 \times 14}$ and ${\bf \hat x}_T \to {\bf \hat X}_T: \mathbb{C}^{1 \times \frac{K_2L}{2}} \to \mathbb{R}^{1\times L \times K_2}$. Then, the signals are semantically recovered information by the channel decoders for text and image, denoted as
\begin{equation}
    {\bf \hat M}_I = {\cal CD}_{I}\left( {\bf \hat X}_{I} ; {{\bm \gamma}_{I}}\right),
\end{equation}
and
\begin{equation}
     {\bf \hat M}_T = {\cal CD}_{T}\left( {\bf \hat  X}_{T} ; {{\bm \gamma}_{T}}\right),
\end{equation}
respectively, where \textcolor{black}{ ${\bf \hat M}_I \in {\mathbb R}^{1\times C_1 \times 14 \times 14}$, ${\bf \hat M}_T \in {\mathbb R}^{1 \times L \times K_1}$,} ${{\bm \gamma}_{I}}$ and ${{\bm \gamma}_{T}}$ are the corresponding trainable parameters. Similar to the channel encoders, the image and text channel decoder consists of CNN layers and dense layers to decompress and recover semantic information.

\begin{figure}[!t]
\centering
\includegraphics[width=85mm]{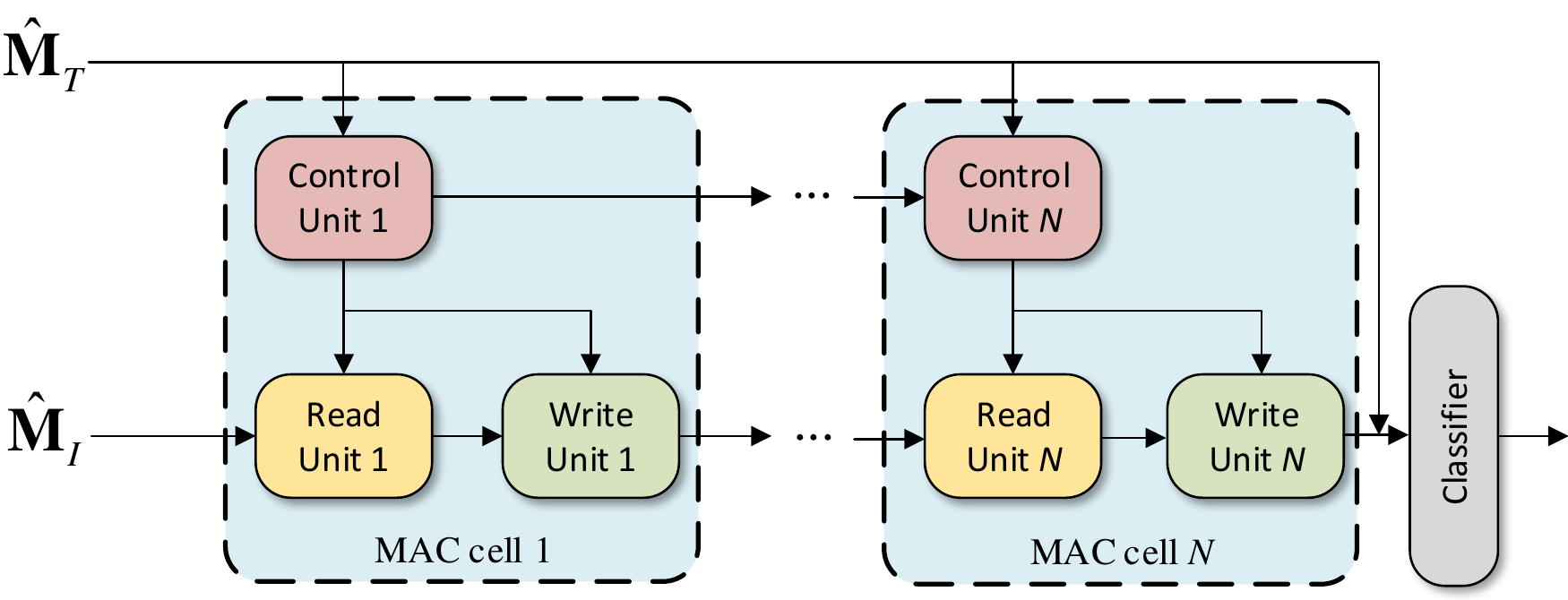}
\caption{\textcolor{black}{The structure of the memory, attention, and composition network for text and image semantic information fusion and the VQA.}}
\label{fig-3}
\end{figure}

With text and image semantic information, we employ the MAC network \cite{HudsonM18} as the semantic decoder to merge the text and image semantic information as well as to answer the vision questions, which is written as
\begin{equation}
    {\tt Task} = {\cal SD}\left( \left ({\bf \hat M}_I, {\bf \hat M}_T \right);{\bm {\varphi}}  \right),
\end{equation}
where  ${\cal SD}\left( \cdot;{\bm {\varphi}}  \right)$ is the semantic decoder with trainable parameters $\bm \varphi$. The MAC network shown in Fig. \ref{fig-3} consists of multiple MAC cells, of which each contains \textit{control unit, read unit}, and \textit{write unit}. The \textit{control unit} first generates a query based on the received text semantic information, i.e., the object of question and the type of question, by the attention mechanism, then the \textit{read unit} gets the query and searches the corresponding key from image semantic information by another attention module. Finally, the \textit{write unit} integrates information and outputs the predicted answers to the questions. 

\subsection{Loss Function}
As indicated before, the objective of the proposed MU-DeepSC is to answer the questions based on the images and texts. The proposed transceiver is task-oriented, where the answers will be predicted directly at the MU-DeepSC receiver. As the image and text will not be recovered in the MU-DeepSC system like traditional communication systems, loss functions based on bit-error or symbol-error are not applicable anymore. In order to improve the accuracy of answers, the cross-entropy (CE) is used as the loss function  to measure the difference between the correct answer, $a$, and the predicted answer, $ \hat a$, which can be formulated as
\begin{equation}\label{loss function 1}
     {\cal L}_{\rm CE}({ a},{\hat a}; {\bm \alpha}, {\bm \beta}, {\bm \gamma}, {\bm \varphi}) =  -  {{p\left( {{a}} \right)}\log \left( {p\left( {\hat a} \right)} \right)},
\end{equation}
where $p(a)$ is the real probability of the answer, and  $p({\hat a})$ is the probability of the predicted answer. The CE can measure the difference between the two probability distributions.  By reducing the loss value of CE, the network learns the correct answer first and tries to predict the answer with the highest probability of accuracy. Then the network can be optimized by gradient descent.  The training procedure is described in Algorithm 1.

\begin{algorithm}[htbp]
\caption{MU-DeepSC Training Algorithm.}\label{algorithm-1}
\small
\SetKwInput{KwInput}{Input}                
\SetKwInput{KwInitia}{Initialization}                
\SetKwInput{KwOutput}{Output}              
\SetKwInput{KwRet}{Return}
\DontPrintSemicolon
  
  \KwInitia{Load pre-trained model for ${{\cal SE}_{ I}\left( {;{{\bm{\alpha }}_I}}\right)}$, and initialize  ${{\bm{\alpha }}_T}, {{\bm{\beta }}_I}, {{\bm{\beta }}_I}, {{\bm{\gamma }}_I}, {{\bm{\gamma }}_T}, {{\bm{\varphi }}}$.}
  \KwData{The training dataset ${\cal D}$.}

  \SetKwFunction{FMain}{Main}
  \SetKwFunction{FSE}{Train Semantic Encoder}
  \SetKwFunction{FCC}{Train Channel Coder}
  \SetKwFunction{FWN}{Train Whole Network}

  \SetKwProg{Fn}{Function}{:}{}
  \Fn{\FWN{}}{
        \textbf{Image Transmitter}:\;
  		\quad $ {{\cal SE}_I\left( {{\bf S}_{I};{{\bm{\alpha }}_I}} \right)}  \to  {\bf M}_{I} $;  $ {{\cal CE}_I\left( {{\bf M}_{I};{{\bm{\beta }}_I}} \right)}  \to  {\bf X}_{I} $, \; 
  		\quad Reshape first ${\bf X}_{I}$ and  power normalization by \eqref{norm-1}, \;
  		\quad Transmit ${\bf x}_{I}$  over the channel.\;

  		\textbf{Text Transmitter}:\;
  		\quad Embedding the input sentence ${\bf s}_T \to {\bf S}_T$,\; 
  		\quad $ {{\cal SE}_T\left( {{\bf S}_T;{{\bm{\alpha }}_T}} \right)}  \to  {\bf M}_T $;  $ {{\cal CE}_T\left( {{\bf M}_T;{{\bm{\beta }}_T}} \right)}  \to  {\bm X}_T $, \; 
  		\quad Reshape ${\bf X}_{T}$ first and power normalization by \eqref{norm-2}, \;
  		\quad Transmit ${\bf x}_{T}$  over the channel.\;

  		\textbf{Receiver}:\;
  		\quad Receive ${\bf Y}$ and Signal detection by \eqref{eq13} to get ${\bf \hat x}_{I}$, ${\bf \hat x}_{T}$,\;
  	    \quad $ {{\cal CD}_I\left( {\hat {\bf x}_I;{{\bm{\gamma }}_I}} \right)} \to  {\hat {\bf M}}_I$; $ {{\cal CD}_T\left( {\hat {\bf x}_T;{{\bm{\gamma }}_T}} \right)} \to  {\hat {\bf M}}_T $, \;
  		\quad $ {{\cal SD}\left( {\left( {\hat {\bf M}}_I, {\hat {\bf M}}_T \right);{{\bm{\varphi }}}} \right)} \to  {\hat a} $, \;
  		Compute the loss by \eqref{loss function 1} with $a$, ${\hat a}$.\;
        Train ${{\bm{\alpha }}_T}, {{\bm{\beta }}_I}, {{\bm{\beta }}_I}, {{\bm{\gamma }}_I}, {{\bm{\gamma }}_T}, {{\bm{\varphi }}}$ $\to$ Gradient descent. \;
        \KwRet{${{\cal SE}_I\left( {;{{\bm{\alpha }}_I}} \right)}, {{\cal SE}_T\left( {;{{\bm{\alpha }}_T}} \right)}, {{\cal CE}_I\left( {;{{\bm{\beta }}_I}} \right)}, {{\cal CE}_T\left( {;{{\bm{\beta }}_T}} \right)}$, ${{\cal CD}_I\left( {;{{\bm{\gamma }}_I}} \right)}$, ${{\cal CD}_T\left( {;{{\bm{\gamma }}_T}} \right)}$, and ${{\cal SD}\left( {;{{\bm{\varphi }}}} \right)}$} 
  }
\end{algorithm}

\section{Simulation Results and Discussions}
In this section, we compare the proposed MU-DeepSC and the traditional source coding and channel coding methods over different channels, where the perfect CSI is assumed for all methods. The transceiver is assumed with two single-antenna users and the receiver with two antennas.

\subsection{Implementation Settings}
The adopted Dataset is CLEVR \cite{johnson2017clevr}, which consists of a training set of 70,000 images and 699,989 questions and a test set of 15,000 images and 149,991 questions.

In the experiments, text will be embedded by embedding layer, which is initialized from Gaussian distribution with zero mean and unit variance, ${\cal N}(0,1)$, with shape (vocab size, embedding-dim). The embedding dimension is set to be 300. $C_1$, $C_2$, $K_1$, and $K_2$ are set 512, 128, 512, and 256, respectively. The image channel coder consists of four convolutional layers with 256, 128, 256, 512 filters, the first two of which are image channel encoder, the rest of which are image channel decoder. Each convolutional layer is with a $3 \times 3$ kernel and followed by an ELU activation function. The text channel coder consists of five dense layers with 256, 256, 256, 256, 512 neurons, the first two of which are text channel encoder, the rest of which are text channel decoder. Each Dense layer is followed by a ReLU activation function and the outputs of the channel decoder are normalized by LayerNorm. The MAC network consists of 12 cells. We employ the ADAM optimizer with a learning rate of 0.0001. Different from predicting with image and text, we mainly consider two cases for the baselines, only using question or image to predict the answer, and the typical separate source and channel coding,
\begin{itemize}
    \item Error-free transmission: The full, noiseless images and texts are input ResNet-101 and Bi-LSTM for extracting features and then input to the MAC network. 
    \item Traditional method: To perform the source and channel coding separately, we use the following technologies respectively:
    \begin{itemize}
        \item Joint photographic experts group (JEPG)  as the image source coding, a commonly used method of lossy compression with a compression rate of 75 for digital images, Huffman coding for text source coding, lossless compression for text, 
        \item Low-density parity-check codes (LDPC) with a coding rate of 1/3  as the channel coding, especially for the large data size. 
    \end{itemize}
    \item Text or image only based prediction: The transmitter is the same as the text user or image user in Fig. \ref{fig-2}, but the receiver replaces the MAC network with a one-layer classifier.
\end{itemize}
The modulation method for the traditional method is 16 quadrature amplitude modulation (QAM).  For the traditional method, the recovered image and text are input to the MAC network to get answers. We compare the proposed methods in terms of answer accuracy, the number of transmitted symbols, and computational complexity.

\subsection{Performance of MU-DeepSC}
\begin{figure*}[htbp]
\centering
\hspace{-9mm}
\subfigure[AWGN Channels.]{
        \centering
        \includegraphics[width=62mm]{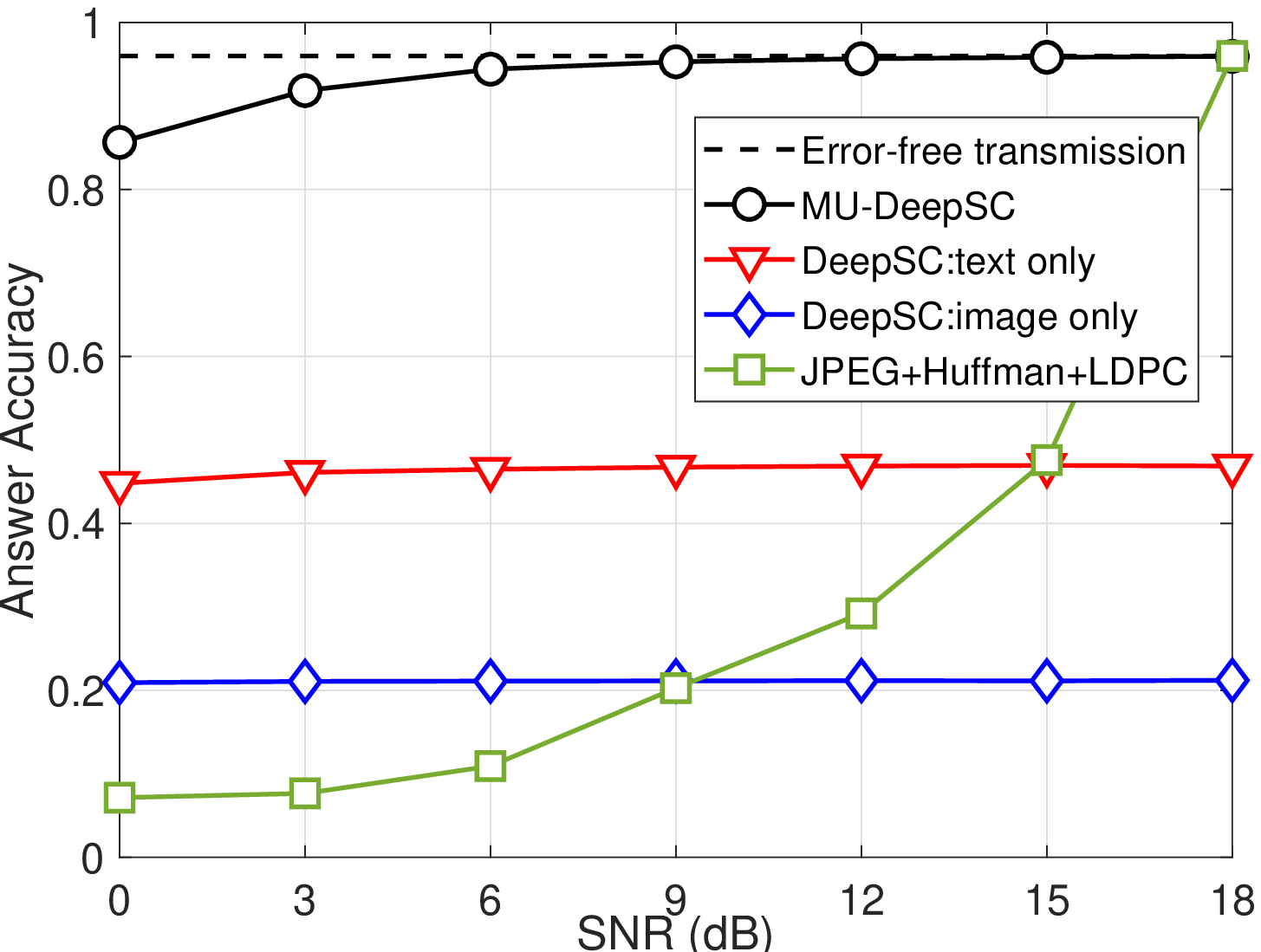}
        \label{fig-4-a}
}\hspace{-3mm}
\subfigure[Rayleigh Fading Channels.]{
        \centering
        \includegraphics[width=62mm]{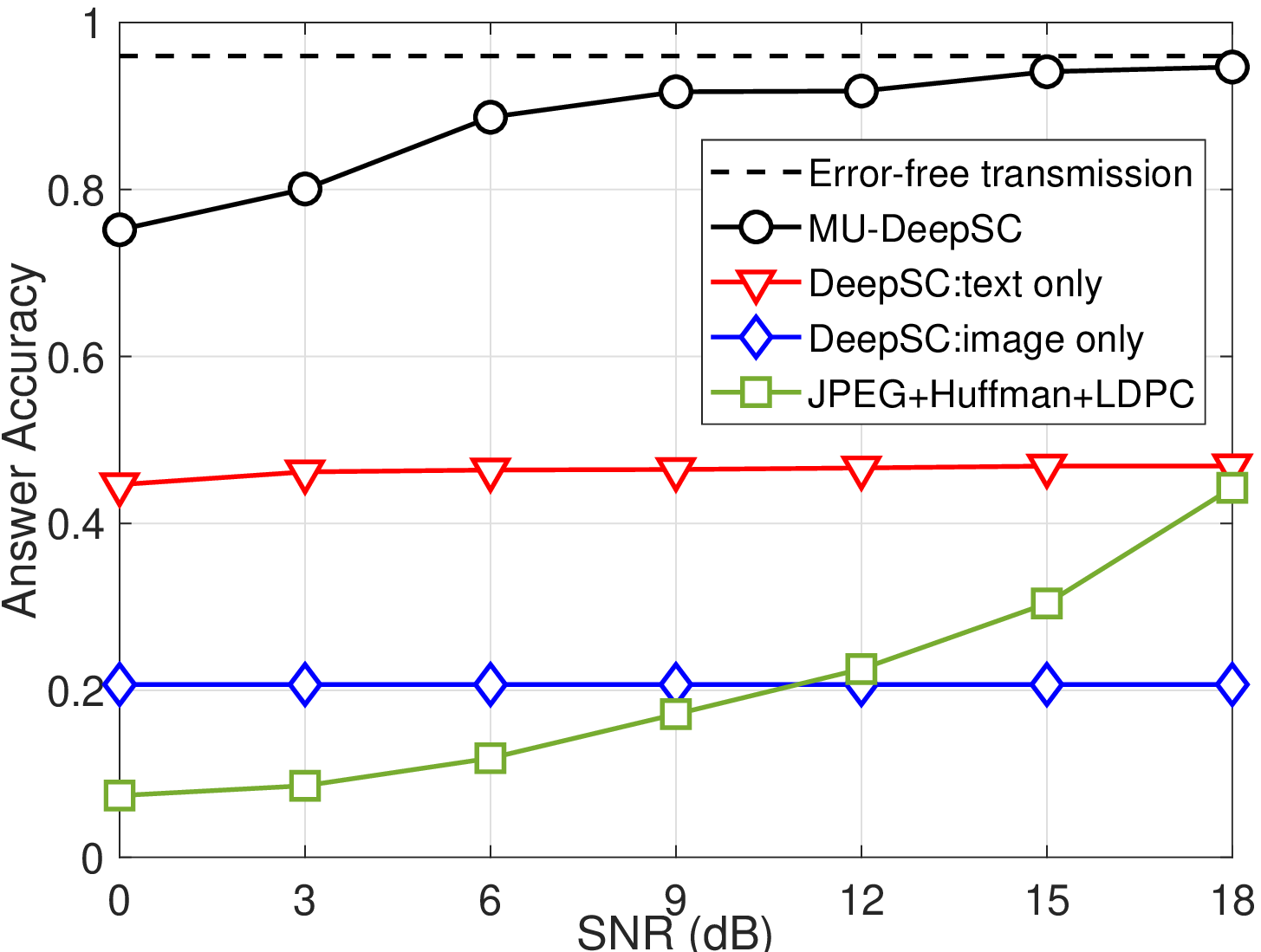}
        \label{fig-4-b}
}\hspace{-3mm}
\subfigure[Rician Fading Channels.]{
        \centering
        \includegraphics[width=62mm]{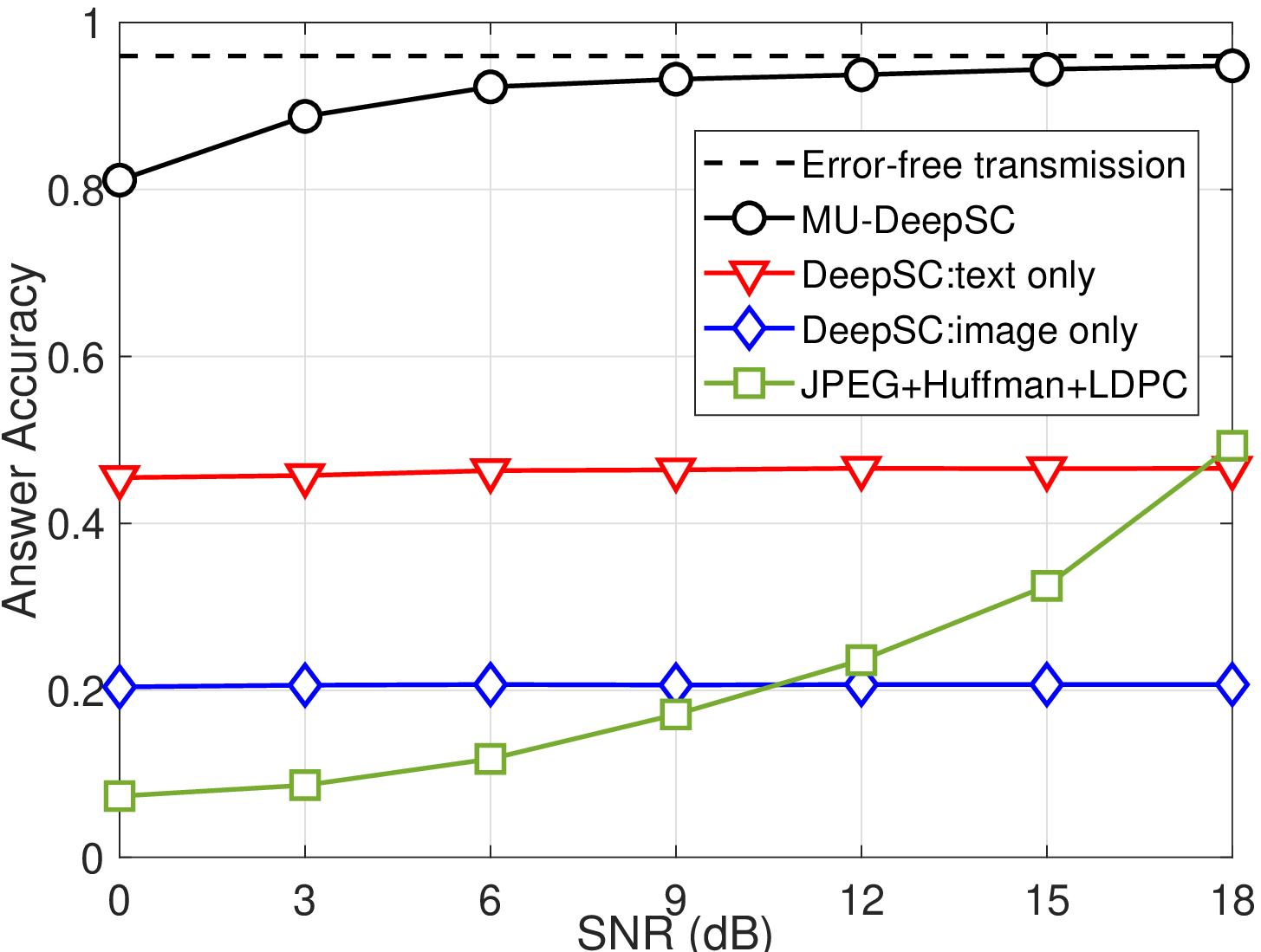}
        \label{fig-4-c}
}\hspace{-9mm}
\caption{Answer accuracy for various testing channels based on different trained models.}
\label{fig-4}
\end{figure*}

Fig. \ref{fig-4} shows the relationship between the answers accuracy and  SNR over AWGN, Rayleigh fading channels, and Rician fading channels. Among the methods in Fig. \ref{fig-4}, the proposed MU-DeepSC outperforms other baselines, especially in the low SNR regime,  and is about to approach the upper bound at high SNR regime.  Besides, over all SNR regimes, transmitting single source, such as text or image only, has the similar answer accuracy over three channels. Moreover, in Fig. \ref{fig-4-a}, the traditional method performs worse at the lower SNR regime since the images are corrupted by error bits, but performing higher accuracy as the SNR increases. For more complex channels, the answer accuracy of the traditional method can increase slowly as SNR increases in Fig. \ref{fig-4-b} and \ref{fig-4-c}.  Besides, compared with the separate source-channel coding in traditional communications, the proposed MU-DeepSC is jointly optimized to achieve better performance at the answer accuracy. 

\begin{figure*}[htbp]
\centering
\includegraphics[width=140mm]{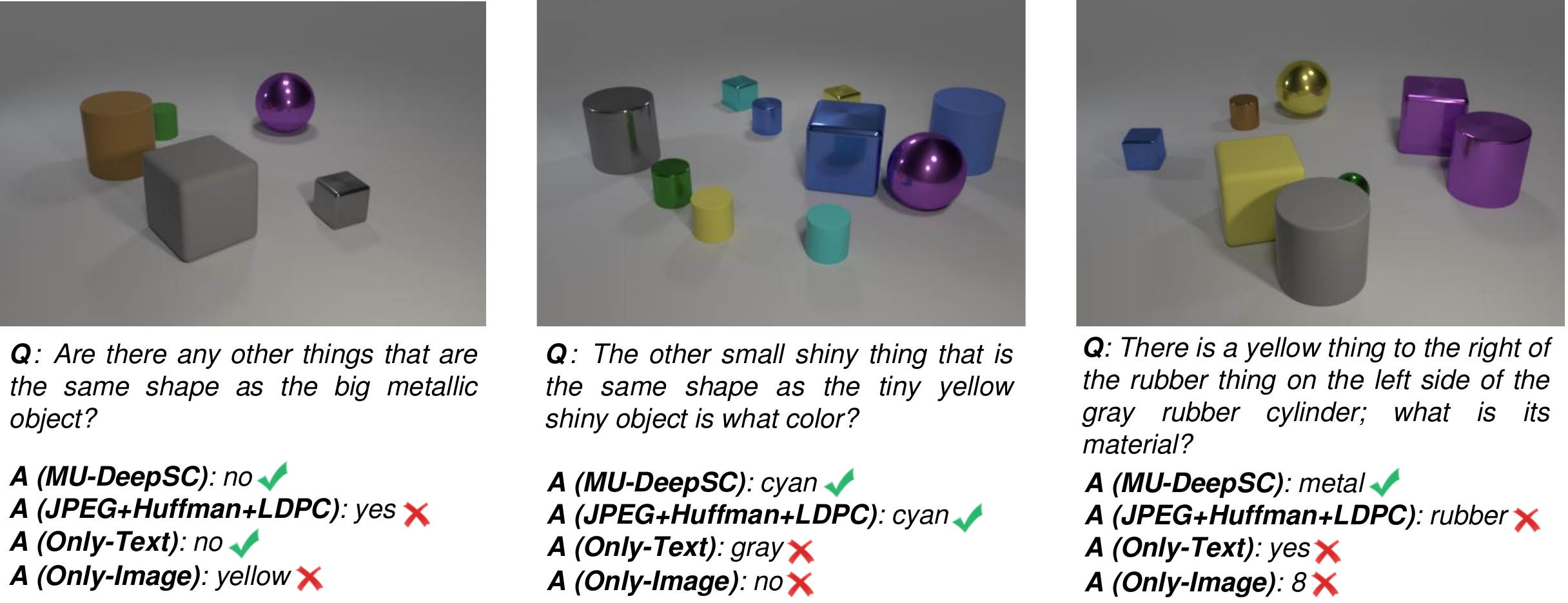}
\caption{Some visualized results for VQA task over Rician fading channels with SNR = 18dB , where the first row is the transmitted images, the second row is the transmitted questions, and the last four rows are predicted answers by proposed MU-DeepSC, traditional methods, MU-DeepSC with text only, and MU-DeepSC with image only, respectively.}
\label{fig-5}
\end{figure*}

Part of visualized results are shown in Fig. \ref{fig-5}. The proposed MU-DeepSC correctly answers the all questions. The traditional communications and the DeepSC based text can only give the right answers for {partial} questions. {The reason that transmitting text-only answers correctly is that the text information can help the system guess and narrow the search range of answers.}

Table \ref{tab:1} compares the proposed MU-DeepSC and traditional communications at the number of transmitted symbols and computational complexity\footnote{We only analyze the complexity of channel coding for both methods because the other parts are shared in both methods and the complexity of source coding is low to be omitted.} by measuring one image or one word. For image transmission, the proposed MU-DeepSC significantly decreases the number of transmitted symbols and the computational complexity. For text transmission, the MU-DeepSC transmits more symbols than the traditional communications but with similar computational complexity, in which a larger number of symbols can provide robustness to channels and low SNRs. In general, the proposed MU-DeepSC can save the transmission and processing time for images, by slightly sacrificing the transmission time for text but keeping a similar processing time. 

\begin{table}[!t]
\caption{Comparison between MU-DeepSC and traditional communications at the number of transmission symbols and computational complexity for one image or one word }
\label{tab:1}
\centering
\scriptsize
\begin{tabular}{ c|c|c|c|c } 
\toprule
\multirow{2}{*}{Method}  & \multirow{2}{*}{Source} & \multirow{2}{*}{No. Symbols} & \multicolumn{2}{c}{Computational Complexity}\\
\cmidrule{4-5}
& & & Additions  &  Multiplications\\
\midrule
\multirow{2}{*}{MU-DeepSC} & Image & 12,544  & $2.6\times 10^8$ & $2.9\times 10^8$\\
\cmidrule{2-5}
& Text & 128  & $4.6 \times 10^5 $ & $4.6 \times 10^5$\\
\midrule
\multirow{2}{*}{\makecell[c]{Traditional \\ Communications}} & Image  & 41,718  & $1.0\times 10^9$ & $1.1\times 10^9$\\
\cmidrule{2-5}
& Text  & 15 & $6.8 \times 10^5 $ &  $3.6 \times 10^5 $\\
\bottomrule
\end{tabular}
\end{table}

\section{Conclusion}
In this letter, we have established a multi-user semantic communication system, named MU-DeepSC, for exploiting the correlated image and text information, where the VQA task is considered. By jointly designing the semantic encoder and the channel encoder by learning and extracting the essential semantic information, the proposed MU-DeepSC can handle the image and text semantic information effectively and predict the answers accurately by merging different semantic information.  The simulation results have demonstrated that the MU-DeepSC outperforms various benchmarks, especially in the low SNR regime. Hence, we are highly confident that the proposed MU-DeepSC is a promising candidate for multi-user semantic communication systems to transmit multimodal data.


%





\ifCLASSOPTIONcaptionsoff
  \newpage
\fi



\vspace{-2.5mm}
\bibliographystyle{IEEEtran}
\bibliography{reference.bib}
\end{document}